\documentstyle[12pt]{article}
\begin{document}
\setlength{\unitlength}{1.0mm}\thicklines
{\hfill Preprint IHEP 94-3\\
        Received 25 January 1994}
\begin{center}
{\bf Scaling law\\ for the
$\Upsilon(4S) \to B \bar B$ and $\psi(3770) \to D \bar D$ decay constants\\
from effective sum rules }\\
\vspace*{1cm}
V.V.Kiselev\\
Institute for High Energy Physics,\\
Protvino, Moscow Region, 142284, Russia.\\
E-mail address: kiselev@mx.ihep.su\\
\vspace*{1cm}
\underline{Abstract}\\
\end{center}

Sum rules for exclusive production of heavy meson pairs in
$e^+e^-$ annihilation are used to evaluate the
$\Upsilon(4S) \to B \bar B$ and $\psi(3770) \to D \bar D$ decay widths.
Infinitely heavy quark limit is discussed, so that scaling law for the
quarkonium-meson coupling constant is derived. A value of the $B\bar B$ pair
contribution into the leptonic constant $f_{\Upsilon(4S)}$ is estimated.
\newpage
\section{Introduction}

Some $\Upsilon$-states, being heavier than a kinematical threshold
of decay into $\bar B B$ pairs, have total widths, which are $10^3$ times
larger than hadronic widths of underthreshold states \cite{1}. The fact
agrees absolutely with the Okubo-Zweig-Iizuka rule. However, concerning
approaces of numerical estimates for the decay widths like $\Gamma(\Upsilon(4S)
\to B\bar B)$, one finds an essential problem, caused by the nonperturbative
character of the calculations, because the strong interaction in the
sector of light quarks, coupling with the heavy $b$-quark into the $B$ meson,
can not be described by the
perturbative theory of QCD. At present, an estimate of the width for the decay
$\psi(3770) \to D \bar D$ was only performed in the framework of the
phenomenological nonrelativistic model \cite{2} by calculating an overlapping
of the meson wave functions. We use another effective way to calculate
nonperturbative quatities, so sume rules are applied \cite{3}.

So, we preliminaraly calculate the total cross sections for the
exclusive production
of the heavy meson pairs in $e^+e^-$ annihilation by use of a simple
constituent quark model, which contains a bit of well defined parameters
\cite{4}. The application of the sum rules allows one to determine the
contribution of the $B \bar B$ pairs into the leptonic decay constant of
$\Upsilon(4S)$. The last value is generally determined by the choice of the
light quark constituent mass in the $B$ meson. The dispersion relation
allows one to calculate the coupling constant of $\Upsilon(4S)$ with the
$B$ mesons. The constant depends slightly on the choice of the model
perameters,
such as the constituent mass of light quark, a threshold of a nonresonant
contribution into the hadronic spectral density.

We study the infinitely heavy quark limit ($m_{b,c} \to \infty$), so that for
the
coupling constant $g$ of the heavy vector quarkonium ($Q\bar Q$) with
the meson ($Q\bar q$), the scaling law is found
$$
\frac{g^2}{M} = const
$$
independently of the heavy quark flavour. This law considers to be rather
reliable for the $\Upsilon(4S) \to B \bar B$ and $\psi(3770) \to D \bar D$
decays.

The model allows one also to consider a possible value of the $m_{light}/
m_{heavy}$ corrections from a phenomenological point of view.

\section{Exclusive Production of Heavy Meson Pair
in $e^+e^-$ Annihilation}

In the framework of the consituent quark model \cite{4} the exclusive
production of the heavy meson pair is determined by the values of the
leptonic decay constant $f$ of the meson, the constituent quark masses
$m_{Q,q}$ and the QCD coupling constant $\alpha_S(4 m_q^2)$. One can
easily show, if $m_Q$ and $m_q$ are masses of the heavy and light quarks,
respectively, then one has for the cross sections of the pseudoscalar $P$
and vector $V$ mesons the following expressions
\begin{eqnarray}
{}~ & ~ &\sigma(e^+e^- \to (Q \bar q)_P (\bar Q q)_P)  =
\frac{\pi^3 \alpha_S^2(4 m_q^2) \alpha^2_{em}}{3^7 \cdot 4 m_q^6}\;
\frac{m_Q^2}{M^2}\; f_P^4 (1-v^2)^3 v^3  \times \nonumber \\
{}~ & ~ &
{}~~~~\biggl(3 e_Q \biggl(\frac{2 m_q}{m_Q} - 1+v^2\biggr)
 - 3e_q \biggl(2 - (1-v^2)\frac{m_q}{m_Q}\biggr)
\frac{m_q^3 \alpha_S(4 m_Q^2)}{m_Q^3 \alpha_S(4 m_q^2)}\biggr)^2\;, \nonumber
\\
{}~ & ~ & \sigma(e^+e^- \to (Q \bar q)_P (\bar Q q)_V)  =
\frac{\pi^3 \alpha_S^2(4 m_q^2) \alpha^2_{em}}{3^7 \cdot 2 m_q^6}\;
f_P^2 f_V^2 (1-v^2)^4 v^3  \times \nonumber \\
{}~ & ~ & ~~~~\biggl(3 e_Q - 3e_q
\frac{m_q^3 \alpha_S(4 m_Q^2)}{m_Q^3 \alpha_S(4 m_q^2)}\biggr)^2\;, \label{1}
\\
{}~ & ~ & \sigma(e^+e^- \to (Q \bar q)_V (\bar Q q)_V)  =
\frac{\pi^3 \alpha_S^2(4 m_q^2) \alpha^2_{em}}{3^7 \cdot 2 m_q^6}\;
f_V^4 (1-v^2)^3 v^3 \times \nonumber \\
{}~ & ~ & ~~~~\biggl(3 e_Q - 3e_q
\frac{m_q^3 \alpha_S(4 m_Q^2)}{m_Q^3 \alpha_S(4 m_q^2)}\biggr)^2
[3 (1-v^2) + (1+v^2)(1-a)^2 + \nonumber \\
{}~ & ~ & ~~~~\frac{a^2}{2} (1-v^2)
(1-3v^2)]\;, \nonumber
\end{eqnarray}
where $v$ is the meson velocity in the center of mass system, and
$$
a = \frac{m_Q}{M}\;\frac
{1-\frac{e_q}{e_Q}\;\frac{m_q^4}{m_Q^4}\;
\frac{\alpha_S(4 m_Q^2)}{\alpha_S(4m_q^2)}}
{1-\frac{e_q}{e_Q}\;\frac{m_q^3}{m_Q^3}\;
\frac{\alpha_S(4 m_Q^2)}{\alpha_S(4m_q^2)}}
$$

One can see in the limit of $m_{light}/m_{heavy} \ll 1$ the values of the cross
sections depend strongly on the choice of the constituent light quark
mass $m_q$.

Note, that near the threshold the cross section ratios for the pseudoscalar
and vector mesons have the form
\begin{equation}
\sigma(PP) : \sigma(PV) : \sigma(VV) = 1 : 4 : 7\;, \label{2}
\end{equation}
where one takes into account $\sigma(PV)=\sigma(B\bar B^*)+\sigma(\bar B B^*)=
2 \sigma(\bar B B^*)$. Eq.(\ref{2}) agrees with EHQT \cite{5} and was obtained
in papers of Refs.\cite{2,6,7}.

\section{$B\bar B$ Contribution into Leptonic Constant $f_{\Upsilon(4S)}$}

In QCD sum rules \cite{3} one found that with required accuracy
\begin{equation}
\sum_n \frac{f_{Vn}^2 M_{Vn}^2}{M_{Vn}^2 - q^2}
\approx \frac{1}{\pi} \int_{s_i}^{s_{th}}
\frac{\Im m \Pi_V^{(QCD)pert}(s)}{s-q^2} ds + \Pi_V^{(QCD)nonpert}(q^2)\;,
\label{3}
\end{equation}
where $f_V$ is leptonic constant of the vector state $V(Q\bar Q)$ with the mass
$M_V$,
\begin{eqnarray}
i f_V M_V \epsilon_\mu^{(\lambda)} e^{ipx} & = &
<0|J_\mu (x)|V(\lambda)>\;, \label{4} \\
J_\mu & = & \bar Q(x) \gamma_\mu Q(x)\;, \nonumber
\end{eqnarray}
and $\Im m \Pi_V^{(QCD)pert}(s)$ is calculated by means of the QCD
perturbation theory, so that
\begin{eqnarray}
{}~ & ~ & (-g_{\mu\nu}+\frac{q_\mu q_\nu}{q^2})\Pi_V^{(QCD)pert}(q^2) +
q_\mu q_\nu \Pi_S^{(QCD)pert}(q^2) = \nonumber \\
{}~ & ~ & ~~~~~~~~~\int d^4x e^{iqx}
<0|T J_\mu(x) J_\nu(0)|0>\;,
\end{eqnarray}
and
\begin{equation}
\Pi_V^{(QCD)nonpert}(q^2) = \sum C_i(q^2) O^i\;,
\end{equation}
where $O^i$ are vacuum expectation values of composit operators like
$<m \bar \psi \psi>$, $<\alpha_s/\pi\;G_{\mu\nu}^2>$ and so on. The Wilson's
coefficients $C_i(q^2)$ are calculated in the perturbative QCD framework.
$s_i$ is the kinematical threshold of the QCD perturbative contribution,
$s_{th}$ is the threshold energy squared for the nonresonant hadronic
contribution, which was kept equal to the perturbative one at $s > s_{th}$.
The sum over the vector resonances in Eq.(\ref{3})
is performed for the states with the
masses $M_{Vn}^2 > s_i$.

It is rather evident that the saturation of the two-point quark current
correlator $<0|T J_\mu(x) J_\nu(0)|0>$ by the two-point meson current
correlator
$<0|T J_\mu^{B \bar B}(x) J_\nu^{B \bar B}(0)|0>$ leads to extraction of the
$B \bar B$ pair contribution into $f_{\Upsilon(4S)}$ for the kinematicaly
admissible lightest state $\Upsilon(4S)$.

One can easily find that
\begin{equation}
\Im m \Pi_V^{B \bar B}(q^2) = \frac{q^4}{(4 \pi \alpha_{em} e_b)^2}\;
\sigma_{e^+e^- \to B \bar B}(q^2)\;.
\end{equation}
Determining the $B$ meson velocity by the expression
\begin{equation}
v = \sqrt{1-\frac{4 M_B^2}{s}}\;, \label{8}
\end{equation}
one states the sum rule for the $B \bar B$ pair contribution into
$f_{\Upsilon(4S)}$
\begin{equation}
\frac{f_{\Upsilon(4S)}^2(B \bar B) M^2_{\Upsilon(4S)}}{M_{\Upsilon(4S)}^2-q^2}
= \frac{1}{\pi}\;\int_0^{v_{th}} \frac{dv^2}{(1-v^2)^3}
\biggl(\frac{M_B^2}{\pi \alpha_{em} e_b}\biggr)^2\; \frac{\sigma_{e^+e^- \to
B \bar B}(v)}{1 - q^2/s}\;, \label{9}
\end{equation}
where one has written down the hadronic part of the sum rule by the two
contributions, so that the first one is the contribution of the $\Upsilon(4S)$
state, lying over the $B \bar B$ production threshold, and the
second contribution is nonresonant one, which coincides with the
calculated cross section of the $B \bar B$ production at $s > s_{th}$.

At $q^2 = 0$  for the n-th derivative of Eq.(\ref{9}) one gets
\begin{equation}
f_{\Upsilon(4S)}^2(B \bar B) = \frac{1}{\pi}\;\int_0^{v_{th}}
\frac{dv^2}{(1-v^2)^{(3-n)}}
\biggl(\frac{M_B^2}{\pi \alpha_{em} e_b}\biggr)^2\; \sigma_{e^+e^- \to
B \bar B}(v)\;, \label{10}
\end{equation}
where in accordance with Eq.(\ref{8}) $v_{th}$ is determined by the threshold
value of the total energy
\begin{equation}
\sqrt{s_{th}} = 2 M_B + \delta\;. \label{100}
\end{equation}
One has  supposed $M_{\Upsilon(4S)} \simeq 2 M_B$.

Some experiences with the QCD sum rule preparation cause to believe that
$s_{th}$ is determined by the energy, at which a production of both
new states and manyparticle systems becomes significant. In the case
interested,
this energy is defined by the threshold of the $B \bar B^*$ and $B^* \bar B^*$
production. The $B^* \bar B^*$ production cross section increases
intensively, beginning from the threshold, so that $\sigma(B^* \bar B^*)/
\sigma(B\bar B) \simeq 7$, and the $B \bar B^*$ production cross section
becomes equal to $\sigma(B \bar B)$ at the same energy of $B^* \bar B^*$
threshold. A new channel opens when additional $\pi$ mesons are
produced.

So, one takes $\delta$ in Eq.(\ref{100}) as
\begin{equation}
\delta \approx 2 (M_{B^*} - M_B) \simeq 100\;\;MeV\;.
\end{equation}
Eq.(\ref{10}) defines the $B \bar B$ pair contribution into leptonic constant
of $\Upsilon(4S)$.

{}From the definition (\ref{4}) it follows that
\begin{equation}
\Gamma(\Upsilon(4S) \to e^+e^-) = \frac{4 \pi}{3}\;\alpha_{em}^2 e_b^2
\; \frac{f^2_{\Upsilon(4S)}}{M_{\Upsilon(4S)}}\;,
\end{equation}
so that one has experimentally \cite{1}
\begin{equation}
f_{\Upsilon(4S)} \simeq 320\;\;MeV\;.
\end{equation}
Taking into account the contributions by the charged and neutral $B$ mesons,
one gets for the  ratio
$$
r = \frac{f_{\Upsilon(4S)}(B \bar B)}{f_{\Upsilon(4S)}}
$$
the following expression
\begin{equation}
r^2 = \frac{4}{\pi f^2_{\Upsilon(4S)}}\; \int_0^{v_{th}}
\frac{dv^2}{(1 -v^2)^{3-n}} \; \biggl(\frac{M_B^2}{\pi \alpha_{em}
e_b}\biggr)^2
\sigma_{e^+e^- \to B \bar B}(v)\;.
\end{equation}
{}From Eq.(\ref{1}) for the $\sigma(e^+e^- \to B \bar B)$ cross section one
can see that the $r$ value depends strongly on the choice of the
constituent light quark mass, so that $\sigma$ is proportional to $1/m_q^6$.

\section{Coupling Constant of $\Upsilon(4S)$ with $B$ Mesons}

Let's consider $f_{\Upsilon(4S)}(B \bar B)$ as a value of a complex
function at the physical point with the condition $\Re e f \gg \Im m f$
near the threshold $q^2 \simeq 4 M_B^2$, where the imaginary part of the
function tends to zero.

Then,  the imaginary part of the transversal correlator, shown on
Fig.1, is related with $\Im m f$ as the following
\begin{equation}
\Im m \Pi^{tr}(q^2) = - M_{\Upsilon(4S)}\; \frac{1}{2}\;\Im m
f_{\Upsilon(4S)}(B \bar B)\;, \label{16}
\end{equation}
so that at $q^2=4 M_B^2/(1-v^2)$ one has
\begin{equation}
\Im m f_{\Upsilon(4S)}(B \bar B) = g_{\Upsilon B \bar B}\;
\frac{\alpha_S}{8\cdot 27} \frac{f_B^2 m_b}{M_B^5}\;
\biggl(\frac{M_B}{m_q}\biggr)^3\;|\vec{k}_B|^3
\;(1-v^2)^2 (1-v^2-\frac{2m_q}{m_b})\;,
\end{equation}
where one has used the formula for the photon-meson vertex, derived in the
framework of the exclusive cross section calculations
\begin{equation}
L_{AB\bar B} = e\cdot e_b\;F(q^2)\;A_\mu\cdot k^\mu\;,
\end{equation}
with
\begin{equation}
F(q^2)= \frac{2 \pi \alpha_S(4 m_q^2)}{9}\;\frac{f_B^2 m_b}{M_B^3}\;
\biggl(\frac{M_B}{m_q}\biggr)^3\;(1-v^2) (1-v^2-\frac{2 m_q}{m_b})\;,
\end{equation}
and the $\Upsilon B \bar B$ vertex has been written down as
\begin{equation}
L_{\Upsilon B \bar B} = g_{\Upsilon B \bar B}\;\epsilon_\mu k^\mu\;,
\label{16a}
\end{equation}
where $\epsilon$ is the polarization vector of $\Upsilon(4S)$,
$k=(p_B - p_{\bar B})/2$.

One can easily find that the consistent consideration of the vector dominance
means that near the threshold
\begin{equation}
(q^2-M_{\Upsilon(4S)}^2)\;\Im m \Pi_V^{B\bar B}(q^2) =
2\;f_{\Upsilon(4S)}(B \bar B) M_{\Upsilon(4S)}\;\Im m \Pi^{tr}(q^2)\;,
\end{equation}
and the sum rules for
$(q^2-M^2_{\Upsilon(4S)})\Pi_V^{B\bar B}(q^2)$ may be rewritten as
\begin{equation}
-f^2_{\Upsilon(4S)}(B \bar B)\;M_{\Upsilon(4S)}^2 =
f_{\Upsilon(4S)}(B \bar B)\;M_{\Upsilon(4S)}\;\frac{2}{\pi}\;
\int_{4M_B^2}^{q^2_{th}} \frac{dq^2}{q^2-s}\;\Im m \Pi^{tr}(q^2)\;,
\end{equation}
so that at physical point $s=4 M_B^2$ one has Eq.(\ref{16}).

Using the dispertion relation like that of Eqs.(\ref{9})-(\ref{10}),
one can easily write down at $q^2=4 M_B^2$
\begin{equation}
\Re e f_{\Upsilon(4S)}(B \bar B) = \frac{1}{\pi}\; \int_0^{v_{th}}
\frac{dv^2}{(1-v^2) v^2}\; \Im m f_{\Upsilon(4S)}(B\bar B)\;.
\end{equation}
Taking into account the condition $|\vec{k}|=v \sqrt{s}/2$,
one obtains
\begin{equation}
\Re e f_{\Upsilon(4S)}(B \bar B) = \frac{1}{\pi}\;\int_0^{v_{th}}
\frac{v^2 dv}{\sqrt{1-v^2}}
g_{\Upsilon B \bar B}\;
\frac{\alpha_S}{4\cdot 27} \frac{f_B^2 m_b}{M_B^2}\;
\biggl(\frac{M_B}{m_q}\biggr)^3\;(1-v^2-\frac{2 m_q}{m_b})\;, \label{18}
\end{equation}
Note, the left hand side of Eq.(\ref{18}) is found from the analogous sum rules
(\ref{9})-(\ref{10}) and the $\Re e f$ value is proportional to $1/m_q^3$,
so that the constant $g_{\Upsilon B \bar B}$ has not a crutial dependence
on the value of the constituent light quark mass.

In accordance with Eq.(\ref{16a}), $g_{\Upsilon B \bar B}$ is related with the
$\Upsilon(4S) \to B^+ B^-$ decay width by the expression
\begin{equation}
\Gamma(\Upsilon(4S) \to B^+ B^-) = \frac{1}{24 \pi} \; g_{\Upsilon B \bar B}^2
\frac{|\vec{k}|^3}{M_{\Upsilon(4S)}^2}\;. \label{19}
\end{equation}
Thus, one can calculate the $\Upsilon(4S) \to B^+ B^-$ decay width by means
of the sum rules. Those calculations may be easily performed explicitly
in the limit of infinitely heavy quark and at $v_{th} \ll 1$.

\section{Infinitely Heavy Quark Limit}

In the limit $m_q/m_b \ll 1$, $m_b \approx M_B$, one can easily find
\begin{equation}
\sigma_{e^+e^- \to B^+ B^-}(v) = \frac{\pi^3}{4\cdot 3^5}\; e_b^2 \alpha_S^2(4
m_q^2) \alpha_{em}^2\;\frac{f_B^4}{m_q^6}\;(1-v^2)^5 v^3\;.
\end{equation}
Then the sum rule (\ref{10}) may be rewritten as
\begin{equation}
f_{\Upsilon(4S)}^2(B^+B^-) = \frac{1}{\pi}\;\int_{0}^{v_{th}}
v^4 dv\;\frac{1}{2\cdot 3^5}\;\alpha_S^2(4 m_q^2)\;\frac{f_B^4 M_B^4}{m_q^6}
(1-v^2)^{2+n}\;.
\end{equation}
At $v_{th} \ll 1$ one gets, that the $B \bar B$ pair contribution into
$f_{\Upsilon(4S)}$ has the form
\begin{equation}
f_{\Upsilon(4S)}(B^+B^-) \approx \frac{\alpha_S(4 m_q^2) f_B^2 M_B^2}
{m_q^3}\;\biggl(\frac{v_{th}^5}{10\cdot 3^5}\biggr)^{1/2}\;. \label{22}
\end{equation}
Furthermore, from the sum rule (\ref{18}) it follows that
\begin{equation}
f_{\Upsilon(4S)}(B^+B^-) \approx g_{\Upsilon B\bar B}\;
\frac{\alpha_S(4 m_q^2)}{4\cdot 81 \pi}\;\frac{f_B^2 M_B^2}
{m_q^3}\;v_{th}^3\;. \label{23}
\end{equation}
{}From Eqs.(\ref{22})-(\ref{23}) one finds
\begin{equation}
g_{\Upsilon B \bar B} \simeq 12 \pi \sqrt{\frac{3}{10 v_{th}}}\;. \label{25}
\end{equation}

Thus, one can see that in the infinitely heavy quark limit the sum rules
for the exclusive production of the $B\bar B$ pairs in the $\Upsilon(4S)$
resonance lead to the expression for $g_{\Upsilon B \bar B}$, which
depends slightly on the threshold of the nonresonant contribution, and the
coupling constant is independent of the parameters like the constituent
light quark mass. One finds conditions like in the effective heavy
quark theory \cite{5}, where, for example, the consideration for the
form factors of semileptonic decays $B \to D^{(*)} e \nu$ leads to the
statement, that the form factor normalization $\xi(q^2=q^2_{max})=1$ is
determined by purely kiematical relations and does not depend on
nonperturbative parameters, such as $m_q$.

\section{Numerical Estimates}

In accordance with the experimental data \cite{1}, Eq.(\ref{19}) allows one to
determine the empirical values of $g_{\Upsilon B \bar B}$ and
$g_{\psi D \bar D}$ for the $\Upsilon(4S)$ and $\psi(3770)$ states,
respectively.

\subsection{Analysis for $\Upsilon(4S)$}

First of all, note that the calculations of the initial five moments of the
spectral function in Eq.(\ref{10}) show that $f_{\Upsilon(4S)}(B \bar B)$
depends slightly on the momentum number $n$, so that $\Delta f/f \approx
5 \%$, i.e. the sum rules give the stable value of $f_{\Upsilon(4S)}(B \bar B)$
under the variation of the scheme parameter ($n$). There is a more essential
dependence on the $f_B$ and $m_q$ choices.

In accordance with the QCD sum rules and the mass spectrum of the mesons with
$b$-quarks, one has supposed \cite{3,77}
\begin{eqnarray}
f_B & = & 100\;\;MeV\;, \nonumber \\
m_q & = & 330\;\;MeV\;, \nonumber \\
m_b & = & 4.95\;\;GeV\;, \\
\delta & = & 100\div 140\;\;MeV\;, \nonumber \\
\Lambda_{QCD} & = & 200\;\;MeV\;. \nonumber
\end{eqnarray}
As for the $m_q/m_b$ correction to the exclusive cross section for the
production of $B \bar B$ pairs, it decreases both the value of
$f_{\Upsilon(4S)}(B \bar B)$ by 15\% and $g_{\Upsilon B \bar B}$ by 10\%,
approximately
\footnote{Note, the value of the first order $1/m_Q$ correction to
the exclusive production of the heavy meson pairs is intentionally
overestimated, since in accordance with the results of Refs.\cite{8a}
the first order correction over $1/m_Q$ must vanish at the threshold,
so that the kinematical correction, following from Eq.(1), must be compensated
by the variation of the leptonic constants due to the $1/m_Q$ correction,
for instance $\Delta f_B/f_B \approx 10\%$.
So, at $v_{th}$ being close to zero,
the corrections to the obtained results are really much less than the
presented estimates, and the results are rather reliable.}.
Thus, one can state that the infinitely heavy quark limit
is valid for the $\Upsilon(4S)$ state with the accuracy about 15\%.

Numerically one has
\begin{eqnarray}
g_{\Upsilon B \bar B}^{exp} & = & 51\pm 3\;, \nonumber \\
g_{\Upsilon B \bar B}^{theor} & = & 53\pm 8\;,  \\
\frac{f_{\Upsilon(4S)}(B \bar B)}{f_{\Upsilon(4S)}} & = & (0.12\div 0.20)\%\;.
\nonumber
\end{eqnarray}

Note, that the main result, concerning the $g$ value, very slightly depends
only on the threshold energy choice ($\delta$), that is reliably fixed by
the physical reason, which is the production of some particles different
from the $\bar B B$ pair, so that the lightest final states are
$\bar B B^*$, $\bar B^* B$, $\bar B^* B^*$.

\subsection{Analysis for $\psi(3770)$}

For the $\psi(3770) \to D\bar D$ decay, the approach accuracy falls
significantly.

First, one observes a strong dependence on the number of the spectral function
moments. For the initial five moments one has $\Delta f/f \approx 30\%$.

Second, the difference between the $m_q/m_b$ corrected and uncorrected
values of $f_{\psi(3770)}(D \bar D)$ or $g_{\psi D\bar D}$ is of the order of
50\% (see the footnote to the previous section).

The variation of the nonresonant contribution threshold is found to be
inessential.

In accordance with the QCD sum rules and the mass spectrum of the mesons with
$c$-quarks, one has supposed \cite{3,77}
\begin{eqnarray}
f_D & = & 170\;\;MeV\;, \nonumber \\
m_q & = & 330\;\;MeV\;, \nonumber \\
m_c & = & 1.54\;\;GeV\;, \\
\delta & = & 200\div 280\;\;MeV\;, \nonumber \\
\Lambda_{QCD} & = & 200\;\;MeV\;. \nonumber
\end{eqnarray}
Then
\begin{eqnarray}
g_{\psi D \bar D}^{exp} & = & 31\pm 2\;, \nonumber \\
g_{\psi D \bar D}^{theor} & = & 30\pm 15\;,  \\
\frac{f_{\psi(3770)}(D \bar D)}{f_{\psi(3770)}} & = & (0.2\div 0.3)\%\;.
\nonumber
\end{eqnarray}

Note, in the limit $m_q/m_{b,c} \to 0$, Eq.(\ref{25}) gives the $g$
values, which are close to the empirical those of both the $\Upsilon$
and $\psi$ particles.

\section{Discussion and Conclusion}

In the present paper we have applied the sum rule approach to analyse
some nonperturbative values such as the $B \bar B$ pair contribution into
the leptonic decay constant of $\Upsilon(4S)$ and the coupling constant of
$\Upsilon(4S)$ with the $B$ mesons.

The consideraton of the infinitely heavy quark limit allows one to derive
a simple explicit expression for $g_{\Upsilon B \bar B}$ independent
of the nonperturbetive parameters like the constituent light quark
mass, so that the $g_{\Upsilon B \bar B}$ value is determined by the
threshold of the nonresonant contribution
$$
g_{\Upsilon B \bar B} \simeq 12 \pi \sqrt{\frac{3}{10 v_{th}}}\;.
$$
Moreover, taking into account that in the EHQT framework the mass
difference for the vector and pseudoscalar states satisfies the relation
\begin{equation}
\Delta M \cdot M = const
\end{equation}
independently of the heavy quark flavour, and
\begin{equation}
v_{th} \simeq \sqrt{2\frac{\Delta M}{M}}\;,
\end{equation}
one gets
\begin{equation}
v_{th} \cdot M = const\;,
\end{equation}
and
\begin{equation}
\frac{g^2}{M} = const \label{33}
\end{equation}
independently of the heavy quark flavour.

The numerical estimates made in accordance with the law (\ref{33}),
are close to the experimental ratio of the $g$ values for $\Upsilon(4S)$
and $\psi(3770)$.

Thus, in the leading order of EHQT the sum rules allow one to state the
scaling law (\ref{33}) for the coupling constants of heavy quarkonia
with heavy mesons, including the single heavy quark\footnote{
Note, that the infinitely heavy quark limit was applied to derive
the scaling law for the leptonic decay constants of heavy quarkonia
($\phi$, $\psi$, $\Upsilon$, $B_c$) in the framework of the QCD sum rules
\cite{8}.

Naive scaling assumptions for the heavy quarkonium were discussed in Refs.
\cite{9}.}.
This law is found to be powerfull for the $\Upsilon(4S)$ and $\psi(3770)$
states. However, it has been shown that the approach accuracy is not high,
when the sum rules are applicated to the $c$-quark systems.
For the $b$-quarks, the
accuracy is rather high and may be considered as admissible.

It has been also shown that the meson pair contribution into the leptonic decay
constants are of the order of 0.2\%.

This work is partially supported by the International Science Foundation
(Soros' programm).

\newpage
\begin{center}
{\large Figure captions}
\end{center}

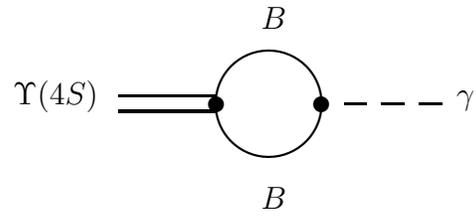
Fig.1. The contribution by the $B$ meson pair into the leptonic decay
constant.
\newpage
\begin{center}
\begin{picture}(90,60)
\put(35,45){\circle{20}}
\put(28,45){\circle*{2}}
\put(42,45){\circle*{2}}
\put(15,44){\line(1,0){13}}
\put(15,46){\line(1,0){13}}

\put(45,45){\line(1,0){3}}
\put(50,45){\line(1,0){3}}
\put(55,45){\line(1,0){3}}

\put(1,45){$\Upsilon(4S)$}
\put(34,31){$B$}
\put(34,55){$B$}
\put(60,45){$\gamma$}
\end{picture}
\end{center}
\begin{center}
Fig.1.
\end{center}
\end{document}